\documentclass[aps,prl,twocolumn,tightenlines,showpacs,aps,amsmath,amssymb,nofootinbib]{revtex4-1}

\usepackage{latexsym}
\usepackage{graphicx}
\usepackage{subfigure}
\usepackage{cases}
\usepackage{hyperref}
\usepackage{amssymb}
\usepackage{bm}
\usepackage{natbib}
\usepackage{amsmath}
\usepackage{slashed}
\usepackage{braket}
\usepackage{cases}

\newcommand{\be}{\begin{equation}}
\newcommand{\ee}{\end{equation}}
\newcommand{\ba}{\begin{eqnarray}}
\newcommand{\ea}{\end{eqnarray}}

\begin{document}

\title{Electromagnetic Currents from Electroweak Fermion Level Crossing}

\author{Eray Sabancilar}
\email[]{Eray.Sabancilar@asu.edu}
\affiliation{Physics Department, Arizona State University, Tempe, Arizona 85287, USA.}


\begin{abstract}
Spectral flow of chiral fermions in the background of electroweak sphalerons is studied. A fermion field configuration that interpolates between the sphaleron zero mode and the asymptotic fermion modes is proposed for the level crossing left handed fermion fields. It is shown that the fermionic electromagnetic currents with non-trivial helicity are produced during the level crossing. Cosmic magnetic field generation due to these currents is briefly discussed. 
\end{abstract}


\maketitle

\emph{Introduction.}---Relics from the early universe, such as, primordial density perturbations, baryon asymmetry, primordial light element abundances, dark matter and cosmic microwave background serve us as extremely powerful tools to study high energy physics and the physics of the early universe. Cosmic magnetic fields generated in the early universe could also have a similar use if they are detected (see, e.g., Ref.~\cite{Durrer:2013pga} for a recent review). For instance, due to an instability in the electroweak plasma \cite{Rubakov:1986am,Rubakov:1985nk}, magnetic fields could be produced if the medium has non-zero right handed electron asymmetry \cite{Joyce:1997uy,Boyarsky:2012ex}. A curious connection between the baryon asymmetry of the universe and the helicity of magnetic fields generated in the electroweak baryogenesis was made \cite{Cornwall:1997ms,Vachaspati:2001nb}, and it was found that this scenario leads to cosmic magnetic fields with left handed helicity \cite{Vachaspati:2001nb,Copi:2008he,Chu:2011tx}. Similarly, in the leptogenesis scenario, right handed magnetic fields are expected to be generated \cite{Long:2013tha}. There is an indirect evidence for the existence of cosmic magnetic fields from TeV blazars \cite{Neronov:1900zz}, and the helicity of such fields could also be determined from the diffuse gamma ray data \cite{Tashiro:2013bxa,Tashiro:2013ita}. Therefore, it is an exciting time to consider the processes in the early universe that can lead to generation of cosmic magnetic fields. In this letter, we will be interested in the question of whether any electromagnetic currents are produced during the level crossing of chiral fermions in the electroweak sphaleron background, which relies only on the standard model physics. We will find that there are indeed non-zero currents with non-trivial helicity, and thus they can lead to cosmic magnetic fields.   


\emph{Topology of Gauge Theories.}---Non-Abelian gauge theories exhibit non-trivial vacuum configuration and topology. This rich structure manifests itself in profound physical processes, such as instantons, sphalerons and level crossing of chiral fermions. Chern-Simons number is a topological invariant of the vacuum manifold of a gauge theory labeling the physically equivalent degenerate vacua. The change in the Chern-Simons number is
\be
\Delta N_{\rm CS} =- \frac{g^2}{32 \pi^2} \int d^4 x ~ {\rm Tr} (F_{\mu \nu} \widetilde F^{\mu\nu}),
\ee
where the gauge connection is $T^a A_{\mu}^a$, the covariant derivative $D_{\mu} = \partial_{\mu} + ig  T^a A_{\mu}^{a}$, the curvature $F_{\mu \nu}  = [D_{\mu},D_{\nu}]$, $\widetilde F^{\mu\nu} = \epsilon^{\mu\nu\alpha\beta} F_{\alpha\beta}/2$ and $g$ is the coupling constant. $\{T^a\}$ are the generators of the gauge group satisfying the Lie algebra $[T^a, T^b] = i f^{abc} T^c$, where $f^{abc}$ are the structure constants [$f^{abc} = \epsilon^{abc}$ for ${\rm SU}(2)$]. Instantons \cite{Belavin:1975fg} and sphalerons \cite{Klinkhamer:1984di,Manton:1983nd} are solutions of the Yang-Mills equations that have non-zero $N_{\rm CS}$, and they exhibit zero modes when coupled to fermions. The Atiyah-Singer index theorem  \cite{Atiyah:1963zz,Atiyah:1968mp} can be used to relate the number of zero modes of the Euclidian Dirac operator to the Chern-Simons number of the gauge fields they are coupled to: ${\rm Index} [\slashed{D}] = \Delta N_{\rm CS}$. The spectral flow of the fermion Hamiltonian is understood as the level crossing of the chiral fermions in the background of topologically non-trivial gauge fields (with non-zero $N_{\rm CS}$). 

The instanton and sphaleron effects lead to the quantum mechanical violation of otherwise classically conserved charges. For instance, the axial charge violating process $\pi^0 \to \gamma \gamma$ occurs due to the triangle diagram that violates this charge \cite{Adler:1969gk,Bell:1969ts}. Similarly, baryon and lepton numbers in the electroweak theory \cite{'tHooft:1976fv,Klinkhamer:1984di} are both anomalous
\be
\partial_{\mu} J^{\mu}_{B} = \partial_{\mu} J^{\mu}_{L} = \frac{g^2}{32 \pi^2} ~ {\rm Tr} (W_{\mu \nu} \widetilde W^{\mu\nu}),
\ee
where  $W_{\mu\nu}^a$ are the weak ${\rm SU(2)}_{\rm L}$ gauge fields. 


\emph{Electroweak Sphaleron.}---In the electroweak theory, the gauge and the Higgs field configurations that interpolate between, e.g., $N_{\rm CS} =1$ and $N_{\rm CS} =0$, have been constructed in Ref.~\cite{Manton:1983nd}:
\ba
T^a A_{i}^{a}  &=& -\frac{i}{g} f(r) (\partial_{i} U) U^{-1}~, \label{A}\\
\Phi &=& [1- h(r) ] \left(\begin{array}{c}0 \\e^{-i\mu} \cos\mu \end{array}\right)_{\tau} + h(r) U \left(\begin{array}{c}0 \\1\end{array}\right)_{\tau},~~ \label{Phi}
\ea
where the ${\rm U(1)}_{Y}$ hypercharge coupling is taken to be $g'\to 0$, $i = \theta, \phi$, $A_{r}^{a} = 0$, $A_{0}^{a}=0$, $T^a \equiv \tau^a/2 = \sigma^a/2$, $\{\sigma^a\}$ are Pauli spin matrices, $f(r)$ and $h(r)$ are radial functions with boundary conditions $f(r)/r|_{r=0} = 0$, $h(0) = 0$, $f(\infty) = h(\infty) = 1$, and are obtained by requiring a finite energy sphaleron solution, and 
\be
U= \left( \begin{array}{cc} e^{i\mu} (\cos\mu -i \sin\mu \cos\theta)  & e^{i \phi} \sin\mu \sin\theta \\
- e^{-i \phi} \sin\mu \sin\theta & e^{-i\mu} (\cos\mu +i \sin\mu \cos\theta) \end{array} \right).
\ee
Here, $\mu \in [0,\pi]$ so that both the gauge fields $A_{i}^{a}$ and the Higgs field $\Phi$ describe the vacuum at $\mu=0$ and $\mu = \pi$, corresponding to $N_{\rm CS} = 0$ and $N_{\rm CS} = 1$, respectively. Using this configuration, one can show that the sphaleron ($\mu = \pi/2$) has $N_{\rm CS} = 1/2$ \cite{Klinkhamer:1984di}. 


\emph{Level Crossing Fermion Configuration.}---Ignoring the fermion masses for simplicity, the Dirac equation for the left handed fermion field reduces to $i \sigma^{\mu} D_{\mu} \Psi_{\rm L} = 0$, or
\be\label{dirac}
i\partial_{t} \Psi_{\rm L} = H_{D} \Psi_{\rm L}.
\ee
Here $\sigma^{\mu} \equiv (I, \boldsymbol{\sigma})$ and the Dirac Hamiltonian is
\be\label{hfull}
H_{D}(\mu) =  \boldsymbol{\sigma} \cdot \left( {\bm p} - g T^a {\bm A}^a \right),
\ee
where $T^a {\bm A}^a$ is given by Eq.~(\ref{A}). At $\mu = \pi/2$ ($N_{\rm CS} = 1/2$) it takes the following form
\be\label{hd}
H_{D}(\pi/2) = \boldsymbol{\sigma} \cdot \hat {\bm r} \left( i\partial_r - \frac{i}{r} \boldsymbol{\sigma} \cdot [{\bm L} + f(r) {\bm T}] \right) - \frac{i}{r} f(r) {\bm T} \cdot \hat {\bm r},
\ee
where $\hat {\bm r}$ is the unit vector along the radial direction, ${\bm L}$ is the angular momentum operator\footnote{Using the identity $ \sigma^i \sigma^j = \delta^{ij} + i \epsilon^{ijk} \sigma^k$ and contracting the indices $i$ and $j$ with position and momentum operators, one gets $ \boldsymbol{\sigma} \cdot {\bm r} ~ \boldsymbol{\sigma} \cdot {\bm p}= {\bm r} \cdot {\bm p} + i \boldsymbol{\sigma} \cdot ({\bm r} \times {\bm p})$. Multiplying this equation from left by $\boldsymbol{\sigma} \cdot {\bm r} $, and using the same identity again, one finally obtains $(\boldsymbol{\sigma} \cdot {\bm r}~ \boldsymbol{\sigma} \cdot {\bm r}) ~\boldsymbol{\sigma} \cdot {\bm p} = r^2 ~\boldsymbol{\sigma} \cdot {\bm p} =\boldsymbol{\sigma} \cdot {\bm r}~ ( {\bm r} \cdot {\bm p} + i \boldsymbol{\sigma} \cdot {\bm L}$).}, whereas the free Hamiltonian is
\be\label{hf}
H_{D}^{\rm free} = \boldsymbol{\sigma} \cdot \hat {\bm r} \left( i\partial_r - \frac{i}{r}[{\bm J}^2 - {\bm L}^2 - {\bm S}^2] \right),
\ee
where we used $ \boldsymbol{\sigma} \cdot {\bm L}  = {\bm J}^2 - {\bm L}^2 - {\bm S}^2$, ${\bm S} \equiv {\boldsymbol \sigma}/2$, ${\bm J} = {\bm L} + {\bm S}$.

It is quite complicated to find all the eigenstates of the full time dependent Hamiltonian since the first and the second term in Eq.~(\ref{hfull}) do not commute in general. Thus, not much progress can be made analytically except in the s-wave sector, where ${\bm L} = 0$. Defining the total angular momentum as ${\bm K} = {\bm J}+{\bm T}$, the eigenvalues of the ${\bm K}^2$ operator are $k=0,1$ for the s-wave sector. The singlet state ($k=0$) is a normalizable zero energy solution, i.e., the fermion zero mode in the electroweak sphaleron background \cite{Ringwald:1988yt}
\be\label{psi0}
\Psi_{\rm sph}^{0}(r) = C_0 ~e^{-2 \int_{0}^{r}dx f(x)/x} ~ \chi,
\ee
where $C_0$ is the normalization constant, $\chi$ is the spin-isospin ($s$-$\tau$) singlet state satisfying ${\bm K}^2 \chi = 0$:
\be\label{chi}
\chi = \frac{1}{\sqrt{2}} \left[ \left(\begin{array}{c}1 \\0\end{array}\right)_{s} \otimes \left(\begin{array}{c}0 \\1\end{array}\right)_{\tau} - \left(\begin{array}{c}0 \\1\end{array}\right)_{s} \otimes \left(\begin{array}{c}1 \\0\end{array}\right)_{\tau}  \right].
\ee 
All the eigenstates of the free Hamiltonian [Eq.~(\ref{hf})] in spherical coordinates can be found analytically for a component of an isospinor:
\be
\Psi_{\ell m}^{\rm free} (r,\theta,\phi)= e^{-i Et} \left[ J_{\ell}(Er)~ \Omega_{\ell m}^{+} - i J_{\ell+1}(Er)~ \Omega_{\ell m}^{-} \right],~~~
\ee
where $\ell$ and $m$ are the eigenvalues of ${\bm L}^2$ and $L_{z}$, respectively, $J_{\ell}(Er)$ is the Bessel function and
\be
\Omega_{\ell m}^{\pm} = \frac{1}{\sqrt{2\ell +1}} \left(\begin{array}{c} \pm \sqrt{\ell \pm m +1/2}~ Y_{\ell}^{m-1/2} (\theta,\phi) \\ \sqrt{\ell \mp m +1/2}~ Y_{\ell}^{m+1/2} (\theta,\phi) \end{array}\right)_{s},
\ee
and $Y_{\ell}^{m \pm 1/2} (\theta, \phi)$ are the spherical harmonics. Note that $\boldsymbol{\sigma} \cdot \hat {\bm r}~\Omega_{\ell m}^{\pm} = -\Omega_{\ell m}^{\mp}$.

Inspecting Eqs.~(\ref{hfull}) and (\ref{hd}) in the limit $f(r) \to 1$, i.e., the pure gauge configuration, 
one can easily see that the spectrum is the same as the free case except that the angular momentum simply shifts, ${\bm L} \to {\bm L}' \equiv {\bm L} + {\bm T}$, hence the eigenstates become $\Psi_{\ell' m'}^{\rm free}$. Therefore, we will argue that the modes with $\ell \neq 0$, do not contribute to the level crossing. Only the singlet state contributes to the level crossing and anomalous particle production\footnote{See, e.g., Ref.~\cite{Diakonov:1993ru} where it was shown numerically that the contribution to the anomalous particle production is due to the singlet state. Also see Refs.~\cite{Gibbons:1993pq,Gibbons:1993hg}, where it was shown analytically that only the singlet state contributes to the level crossing in the gravitational sphaleron background in the manifold $\mathbb{R} \times S^3$.}. 

The full time dependent problem cannot be solved analytically since the Hamiltonian [Eq.~(\ref{hfull})] cannot be diagonalized, however it has been investigated in various numerical studies \cite{Kunz:1993ir,Diakonov:1993ru}. Here we propose a simple fermion configuration so that we can describe the wavefunction of a level crossing massless left handed electroweak fermion doublet analytically as the sphaleron interpolates between the gauge configurations with $N_{\rm CS} = 1$ and $N_{\rm CS} = 0$:
\ba\label{ansatz}
\Psi_{\rm L}({\bm x},t) &=& C(t)~ e^{-i \int_{0}^t dt' E(t')}\nonumber\\
\qquad &\times&~e^{-[8 \mu(t) \overline{\mu}(t)/\pi^2] \int_{0}^{r}dx f(x)/x} ~ \chi.
\ea
Here $E(t) \equiv  E_{0} [1-2\mu(t)/\pi]$ 
is the energy of the fermion state crossing zero, $E_{0}$ is the energy gap between the modes, $\int d^3 x~ |C(t=\pm \infty)|^2 =1$, $C(t=0)=C_0$. For practical purposes we shall use \cite{Chu:2011tx},
\ba
\mu(t) = \pi [1 - \tanh (2t/r_s)]/2,\\
\overline{\mu}(t) = \pi [1 + \tanh (2t/r_s)]/2,
\ea
where $t$ is the physical time and $r_s$ is the sphaleron radius. Note that $\mu(t=-\infty) = \pi$, $\mu(t=0) = \pi/2$ and $\mu(t=\infty) = 0$ so that the fermion energy becomes $E(t= -\infty) = - E_0$, $E(t=0) = 0$ and $E(t=\infty) = E_0$. Similarly, $\overline{\mu}(t=-\infty) = 0$, $\overline{\mu}(t=0) = \pi/2$ and $\overline{\mu}(t=\infty) = \pi$ so that  $\Psi_{\rm L}({\bm x},t \to -\infty) \propto e^{i E_{0} t }~ \chi $, $\Psi_{\rm L}({\bm x},t =0) = \Psi_{\rm sph}^{0}(r) $ and $\Psi_{\rm L}({\bm x},t \to \infty) \propto e^{-i E_{0} t } ~\chi$. The reason why we just focus on the singlet $s$-$\tau$ state $\chi$ given by Eq.~(\ref{chi}), but not higher angular momentum states, is that only the singlet state crosses zero, and the energy of the states with non-zero angular momentum simply shift as we have just argued. Therefore, the proposed configuration given by Eq.~(\ref{ansatz}) describes the fermion wavefunction that smoothly interpolates between the sphaleron zero mode $\Psi_{\rm sph}^{0}(r)$ and the asymptotic singlet states $\Psi_{\rm L}({\bm x},t \to \pm \infty)$, i.e.,  the mode corresponding to the anomalous particle production. 


\emph{Electromagnetic Current.}---The electromagnetic current is usually calculated in the unitary gauge, where
\be
\Phi_{\rm unitary} = \left(\begin{array}{c}0 \\1\end{array}\right),
\ee
however, the Higgs field given by Eq.~(\ref{Phi}) is in a different gauge. We can define the electromagnetic charge operator $Q$ in an arbitrary gauge as \cite{Achucarro:1999it}
\be\label{q}
Q = n^a T^a +\frac{Y_L}{2}, 
\ee
where
\be\label{na}
n^a = - \frac{\Phi^\dagger \tau^a \Phi}{\Phi^\dagger \Phi}.
\ee
$\Phi$ given by Eq.~(\ref{Phi}) can be re-written in the form
\be
\Phi = \left(\begin{array}{c}h(r) e^{i \phi} \sin\mu \sin\theta \\ e^{-i\mu} \left[ \cos\mu +i h(r) \sin\mu \cos\theta \right] \end{array}\right).
\ee
Then, the unit vectors $n^a$ can be found explicitly as
\ba
n^1 &=& - \frac{2 h(r) \sin\mu \sin\theta}{\cos^2\mu + h(r)^2 \sin^2\mu} \biggl[ \cos\mu \cos(\phi+\mu) \nonumber\\ 
	&+& h(r) \sin\mu \cos\theta \sin(\phi+\mu) \biggr], \label{n1}\\
n^2 &=& \frac{2 h(r) \sin\mu \sin\theta}{\cos^2\mu + h(r)^2 \sin^2\mu} \biggl[ \cos\mu \sin(\phi+\mu) \nonumber \\
	&-& h(r) \sin\mu \cos\theta \cos(\phi+\mu) \biggr], \label{n2}\\
n^3 &=& \frac{h(r)^2 \sin^2\mu (\cos^2\theta- \sin^2\theta) + \cos^2\mu}{\cos^2\mu + h(r)^2 \sin^2\mu}. \label{n3}
\ea
Thus, the electromagnetic current density is 
\be
j_{\rm em}^{\nu} = e {\overline \Psi}_{L} (\sigma^{\nu} \otimes Q) \Psi_{L},
\ee
where $e$ is the positron charge. More explicitly,
\be
j_{\rm em}^{\nu}  = \frac{1}{2} j_{Y_{L}}^{\nu}+n^1~  j_{W^{1}}^{\nu} + n^2 ~j_{W^{2}}^{\nu} + n^3 ~j_{W^{3}}^{\nu}, 
\ee
where 
\ba
j_{Y_{L}}^{\nu} &\equiv& e ~Y_{L}~ {\overline \Psi}_{L} (\sigma^{\nu} \otimes I) \Psi_{L},\\
j_{W^{a}}^{\nu} &\equiv& \frac{1}{2} e ~{\overline \Psi}_{L} (\sigma^{\nu} \otimes \tau^a) \Psi_{L}.
\ea
Using Eq.~(\ref{ansatz}), we obtain the contribution from ${\rm U(1)}_{Y}$ current density as
\ba
j_{Y_{L}}^{0} &=& e ~Y_{L}~ {\overline \Psi}_{L} \Psi_{L} = e ~Y_{L} ~ j(r,t), \label{j0y}\\
j_{Y_{L}}^{i} &=& \frac{1}{2} e ~Y_{L}~ j(r,t)~{\rm tr}(\sigma^{i}) = 0, \label{jy}
\ea
where $j(r,t) \equiv |C(t)|^2~ e^{-[16 \mu(t) \overline{\mu}(t)/\pi^2] \int_{0}^{r}dx f(x)/x}$, and we have used ${\rm tr}(\sigma^{i}) = 0$.

Similarly, the contribution from the ${\rm SU(2)}_{L}$ currents are obtained as
\ba
j_{W^{1}}^{0} &=& \frac{1}{2} e ~j(r,t) {\overline \chi} (I \otimes \tau^1) \chi = 0, \label{j01}\\
j_{W^{2}}^{0} &=& \frac{1}{2} e ~j(r,t)~ {\overline \chi}  (I \otimes \tau^2) \chi = 0, \label{j02}\\
j_{W^{3}}^{0} &=&\frac{1}{2} e ~j(r,t)~ {\overline \chi}  (I \otimes \tau^3) \chi = 0, \label{j03}
\ea
and
\ba
{\bm j}_{W^{1}} &=& \frac{e}{2} j(r,t) (-\sin\theta \cos\phi~ \hat {\bf r} - \cos\theta \cos\phi~ \hat {\boldsymbol \theta} + \sin\phi ~\hat {\boldsymbol \phi}), \label{j1}~~~~~\\
{\bm j}_{W^{2}} &=& \frac{e}{2} j(r,t) (-\sin\theta \sin\phi ~\hat {\bf r} - \cos\theta \sin\phi~ \hat {\boldsymbol \theta} - \cos\phi ~\hat {\boldsymbol \phi}),\label{j2}~~~~\\
{\bm j}_{W^{3}} &=& \frac{e}{2} j(r,t) (-\cos\theta~ \hat {\bf r} + \sin\theta ~\hat {\boldsymbol \theta}), \label{j3}
\ea
where $\hat {\bf r}, \hat {\boldsymbol \theta}, \hat {\boldsymbol \phi}$ are the unit vectors in spherical coordinates.

To summarize, we found that during the level crossing the electromagnetic charge and current densities are 
\ba
j_{\rm em}^{0} &=& \frac{1}{2}~ j_{Y_{L}}^{0}, \label{jem0}\\
{\bm j}_{\rm em}  &=& n^1~ {\bm j}_{W^{1}} + n^2 ~{\bm j}_{W^{2}} + n^3 ~{\bm j}_{W^{3}}. \label{jem}
\ea
where $j_{Y_{L}}^{0}$ is given by Eq.~(\ref{j0y}), $\{n^a\}$ are given by Eqs.~(\ref{n1})-(\ref{n3}) and $\{{\bm j}_{W^{i}}\}$ are given by Eqs.~(\ref{j1})-(\ref{j3}).

Note that for each family the quark doublet has hypercharge $Y_{Q} = 1/3$, and the lepton doublet has hypercharge $Y_{L} = -1$. Since the electromagnetic charge density for each doublet [Eq.~(\ref{jem0})] is proportional to the hypercharge, the total charge density vanishes since $2 N_c Y_{Q} + 2 Y_{L} = 0$, where the factor of $2$ accounts for the isospin and $N_c = 3$ accounts for the color degree of freedom of quarks. The electromagnetic current for each quark or lepton doublet is the same since it only gets contribution from the ${\rm SU(2)}_{L}$ sector [Eq.~(\ref{jem})]. Hence, the total current per generation is simply ${\bm j}_{\rm em}^{\rm tot} = (N_c +1) {\bm j}_{\rm em}$, where $N_c=3$ is the color degree of freedom of the quark doublet, and the factor of $1$ accounts for the lepton doublet. Note also that at $\mu = 0, \pi$ ($t \to \pm \infty$) the components of the unit vector $n^a$ [Eq.~(\ref{na})] becomes $n^1 =0$, $n_2 = 0$, $n^3 =1$. The electromagnetic current in this regime goes to ${\bm j}_{\rm em} \to (e/2) (-\cos\theta~ \hat {\bf r} + \sin\theta ~\hat {\boldsymbol \theta})$, thus $\nabla \cdot {\bm j}_{\rm em} \to 0$, as it should.    

\begin{figure}[h]
\includegraphics[scale=0.9]{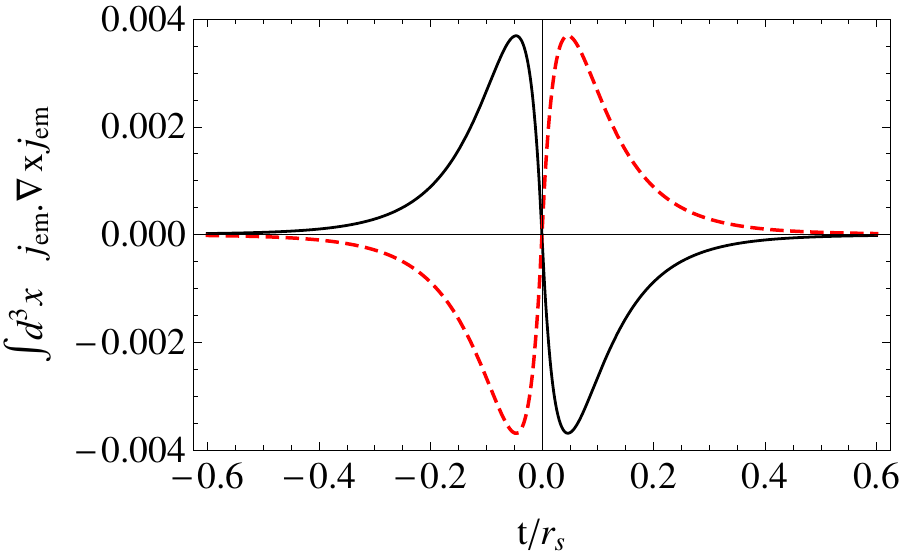}
\caption{$ \int d^{3}x~{\bm j_{\rm em}} \cdot \nabla \times {\bm j_{\rm em}} $ as a function of time in units of $r_s$. Solid (black) line corresponds to transition from $N_{\rm CS}: 1 \to 0$ whereas dotted (red) line corresponds to $N_{\rm CS}: 0 \to 1$.
}
\label{fig:helj}
\end{figure}

For our estimate of the helicity of the current, we use the following approximate profile functions for the gauge and the Higgs fields \cite{Manton:1983nd}:
\begin{numcases}
{f(r) =}
(r/r_s)^2~, \nonumber &
\qquad for $r \leqslant r_s$,
\\
1,  & \qquad for $r > r_s$,
\end{numcases}
\begin{numcases}
{h(r) =}
r/r_s~,  \nonumber&
\qquad for $r \leqslant r_s$,
\\
1,  & \qquad for $r > r_s$.
\end{numcases}
The helicity of the electromagnetic current 
\be
\int d^{3}x~{\bm j_{\rm em}} \cdot \nabla \times {\bm j_{\rm em}},
\ee
is evaluated as can be seen in Fig.~\ref{fig:helj}. Note that the helicity of the current starts out as zero initially, and becomes positive as the level crossing occurs. It becomes zero at the zero mode, and finally changes sign and goes to zero again. This turning on and off of the current helicity can be a hint for generation of cosmic magnetic fields with net helicity.

The peak magnetic field amplitude at the production epoch due to the current given by Eq.~\ref{jem} can be estimated as $B \sim |{\bm j_{\rm em}}|/r \sim e |C_0|^2 / (2 r_s )$, where $|C_0|^2 \approx 0.17/ r_s^3$, hence the maximum magnetic field strength is $B \sim  10^{12}~{\rm G}$, where we used $r_s \sim (100~{\rm GeV})^{-1}$. To estimate the spectrum, helicity and coherence length of these fields, one needs to solve the full magnetohydrodynamic equations taking the plasma effects into account.

In this letter, we proposed a fermion configuration to describe the level crossing of the left handed electroweak fermion doublets in the background of electroweak sphalerons. We showed that fermionic ${\rm U(1)}_{\rm em}$ currents [Eq.~(\ref{jem})] are produced during the level crossing. These currents are also shown to be linked, i.e., they establish non-zero helicity during the level crossing, which could be a hint for magnetic fields with net helicity. There is also a bosonic contribution to the electromagnetic current from the hypercharge current of the Higgs field during the sphaleron transition \cite{Chu:2011tx}. Both fermionic and bosonic contributions should be taken into account to estimate the net magnetic field spectrum from the electroweak sphaleron processes in the early universe. These fields could be the required seed fields to explain the observed galactic magnetic fields, and could be detected in the extragalactic regions by using high energy gamma ray sources \cite{Neronov:1900zz,Tashiro:2013bxa,Tashiro:2013ita}. Observation of these magnetic fields can serve as a handle on the very novel processes, such as, sphalerons, level crossing and anomalous particle production in the early universe. 


I would like to especially thank Tanmay Vachaspati for several discussions and for critical comments. I would also like to thank Andrei Belitsky, Andrew J. Long and Oleg Ruchayskiy for useful discussions, and Howard Haber for pointing out the normalization issue in the charge operator. This work is supported by the National Science Foundation grant No. PHY-1205745 and by the Department of Energy at Arizona State University. 



\begin{thebibliography}{999}

\bibitem{Durrer:2013pga} 
  R.~Durrer and A.~Neronov,
``{\it Cosmological Magnetic Fields: Their Generation, Evolution and Observation}",
 \href{http://arxiv.org/abs/1303.7121}{[arXiv:1303.7121 [astro-ph.CO]]}.

\bibitem{Rubakov:1986am} 
  V.~A.~Rubakov and A.~N.~Tavkhelidze,
``{\it Stable Anomalous States of Superdense Matter in Gauge Theories}",
  Phys.\ Lett.\ B {\bf 165}, 109 (1985).

\bibitem{Rubakov:1985nk} 
  V.~A.~Rubakov,
  ``{\it On the Electroweak Theory at High Fermion Density}",
  Prog.\ Theor.\ Phys.\  {\bf 75}, 366 (1986).
  
\bibitem{Joyce:1997uy} 
  M.~Joyce and M.~E.~Shaposhnikov,
``{\it Primordial magnetic fields, right-handed electrons, and the Abelian anomaly}",
  Phys.\ Rev.\ Lett.\  {\bf 79}, 1193 (1997)
 \href{http://arxiv.org/abs/astro-ph/9703005}{[arXiv: astro-ph/9703005]}.

\bibitem{Boyarsky:2012ex} 
  A.~Boyarsky, O.~Ruchayskiy and M.~Shaposhnikov,
 ``{\it Long-range magnetic fields in the ground state of the Standard Model plasma}",
  Phys.\ Rev.\ Lett.\  {\bf 109}, 111602 (2012)
 \href{http://arxiv.org/abs/1204.3604}{[arXiv:1204.3604 [hep-ph]]}.
  
\bibitem{Cornwall:1997ms} 
  J.~M.~Cornwall,
``{\it Speculations on primordial magnetic helicity}",
  Phys.\ Rev.\ D {\bf 56}, 6146 (1997)
  \href{http://arxiv.org/abs/hep-th/9704022}{[arXiv:hep-th/9704022]}.

\bibitem{Vachaspati:2001nb} 
  T.~Vachaspati,
 ``{\it Estimate of the primordial magnetic field helicity}",
  Phys.\ Rev.\ Lett.\  {\bf 87}, 251302 (2001)
 \href{http://arxiv.org/abs/astro-ph/0101261}{[arXiv:astro-ph/0101261]}.

\bibitem{Copi:2008he} 
  C.~J.~Copi, F.~Ferrer, T.~Vachaspati and A.~Achucarro,
  ``{\it Helical Magnetic Fields from Sphaleron Decay and Baryogenesis}",
  Phys.\ Rev.\ Lett.\  {\bf 101}, 171302 (2008)
  \href{http://arxiv.org/abs/0801.3653}{[arXiv:0801.3653 [astro-ph]]}.

\bibitem{Chu:2011tx} 
  Y.~-Z.~Chu, J.~B.~Dent and T.~Vachaspati,
 ``{\it Magnetic Helicity in Sphaleron Debris}",
  Phys.\ Rev.\ D {\bf 83}, 123530 (2011)
  \href{http://arxiv.org/abs/1105.3744}{[arXiv:1105.3744 [hep-th]]}.

\bibitem{Long:2013tha} 
  A.~J.~Long, E.~Sabancilar and T.~Vachaspati,
 ``{\it Leptogenesis and Primordial Magnetic Fields}",
  \href{http://arxiv.org/abs/1309.2315}{[arXiv:1309.2315 [astro-ph.CO]]}.

\bibitem{Neronov:1900zz} 
  A.~Neronov and I.~Vovk,
 ``{\it Evidence for strong extragalactic magnetic fields from Fermi observations of TeV blazars}",
  Science {\bf 328}, 73 (2010)
  \href{http://arxiv.org/abs/1006.3504}{[arXiv:1006.3504 [astro-ph.HE]]}.

\bibitem{Tashiro:2013bxa} 
  H.~Tashiro and T.~Vachaspati,
  ``{\it Cosmological magnetic field correlators from blazar induced cascade}",
  Phys.\ Rev.\ D {\bf 87}, 123527 (2013)
  \href{http://arxiv.org/abs/1305.0181}{[arXiv:1305.0181 [astro-ph.CO]]}.

\bibitem{Tashiro:2013ita} 
  H.~Tashiro, W.~Chen, F.~Ferrer and T.~Vachaspati,
 ``{\it Search for CP Violation in the Gamma Ray Sky}",
  \href{http://arxiv.org/abs/1310.4826}{[arXiv:1310.4826 [astro-ph.CO]]}.

\bibitem{Belavin:1975fg}
  A.~A.~Belavin, A.~M.~Polyakov, A.~S.~Schwartz and Y.~.S.~Tyupkin,
 ``{\it Pseudoparticle Solutions of the Yang-Mills Equations}",
  Phys.\ Lett.\ B {\bf 59}, 85 (1975).

\bibitem{Manton:1983nd} 
  N.~S.~Manton,
  ``{\it Topology in the Weinberg-Salam Theory}",
  Phys.\ Rev.\ D {\bf 28}, 2019 (1983).

\bibitem{Klinkhamer:1984di} 
  F.~R.~Klinkhamer and N.~S.~Manton,
  ``{\it A Saddle Point Solution in the Weinberg-Salam Theory}",
  Phys.\ Rev.\ D {\bf 30}, 2212 (1984).
  
\bibitem{Atiyah:1963zz} 
  M.~F.~Atiyah and I.~M.~Singer, ``{\it The index of elliptic operators on compact manifolds}",
  Bull.\ Am.\ Math.\ Soc.\  {\bf 69}, 422 (1969).
  
\bibitem{Atiyah:1968mp} 
  M.~F.~Atiyah and I.~M.~Singer,
 ``{\it The Index of elliptic operators. 1}~",
  Annals Math.\  {\bf 87}, 484 (1968).

\bibitem{Adler:1969gk} 
  S.~L.~Adler,
  ``{\it Axial vector vertex in spinor electrodynamics}",
  Phys.\ Rev.\  {\bf 177}, 2426 (1969).

\bibitem{Bell:1969ts} 
  J.~S.~Bell and R.~Jackiw,
 ``{\it A PCAC puzzle: pi0 $\to$ gamma gamma in the sigma model}~",
  Nuovo Cim.\ A {\bf 60}, 47 (1969).

\bibitem{'tHooft:1976fv} 
  G.~'t Hooft,
  ``{\it Computation of the Quantum Effects Due to a Four-Dimensional Pseudoparticle}",
  Phys.\ Rev.\ D {\bf 14}, 3432 (1976)
  [Erratum-ibid.\ D {\bf 18}, 2199 (1978)].
  
\bibitem{Ringwald:1988yt} 
  A.~Ringwald,
  ``{\it Sphaleron And Level Crossing}",
  Phys.\ Lett.\ B {\bf 213}, 61 (1988).
  
\bibitem{Kunz:1993ir} 
  J.~Kunz and Y.~Brihaye,
  ``{\it Fermions in the background of the sphaleron barrier}",
  Phys.\ Lett.\ B {\bf 304}, 141 (1993)
  \href{http://arxiv.org/abs/hep-ph/9302313}{[arXiv:hep-ph/9302313]}.
  
\bibitem{Diakonov:1993ru} 
  D.~Diakonov, M.~V.~Polyakov, P.~Sieber, J.~Schaldach and K.~Goeke,
  ``{\it Fermion sea along the sphaleron barrier}",
  Phys.\ Rev.\ D {\bf 49}, 6864 (1994)
  \href{http://arxiv.org/abs/hep-ph/9311374}{[arXiv:hep-ph/9311374]}.

\bibitem{Gibbons:1993pq} 
  G.~W.~Gibbons and A.~R.~Steif,
  ``{\it Yang-Mills cosmologies and collapsing gravitational sphalerons}",
  Phys.\ Lett.\ B {\bf 320}, 245 (1994)
 \href{http://arxiv.org/abs/hep-th/9311098}{[arXiv:hep-th/9311098]}.

\bibitem{Gibbons:1993hg} 
  G.~W.~Gibbons and A.~R.~Steif,
  ``{\it Anomalous fermion production in gravitational collapse}",
  Phys.\ Lett.\ B {\bf 314}, 13 (1993)
  \href{http://arxiv.org/abs/gr-qc/9305018}{[arXiv:gr-qc/9305018]}.
  
\bibitem{Achucarro:1999it} 
  A.~Achucarro and T.~Vachaspati,
  ``{\it Semilocal and electroweak strings}",
  Phys.\ Rept.\  {\bf 327}, 347 (2000)
  \href{http://arxiv.org/abs/hep-ph/9904229}{[arXiv:hep-ph/9904229]}.
  
\end{thebibliography}
\end{document}